\documentclass[twocolumn,amsmath,amssymb,superscriptaddress,aps,pra,a4paper,floatfix]{revtex4-2}
\UseRawInputEncoding

\usepackage{graphicx}
\usepackage[paperwidth=212mm,paperheight=297mm,centering,hmargin=1.65cm,vmargin=2.6cm]{geometry}
\usepackage{color}
\usepackage{dcolumn}
\usepackage{bm}
\usepackage{physics}
\usepackage{float}
\usepackage{natbib}
\usepackage{amsmath}
\usepackage{xr}
\begin{document}

\title{Entanglement-inspired frequency-agile rangefinding}

\author{Weijie~Nie}
\email{weijie.nie@bristol.ac.uk}
\affiliation{Quantum Engineering Technology Labs, School of Electrical, Electronic and Mechanical Engineering, University of Bristol, BS8 1FD, United Kingdom}
\author{Peide~Zhang}
\affiliation{Quantum Engineering Technology Labs, School of Electrical, Electronic and Mechanical Engineering, University of Bristol, BS8 1FD, United Kingdom}
\author{Alex~McMillan}
\affiliation{Quantum Engineering Technology Labs, School of Electrical, Electronic and Mechanical Engineering, University of Bristol, BS8 1FD, United Kingdom}
\author{Alex~S.~Clark}
\affiliation{Quantum Engineering Technology Labs, School of Electrical, Electronic and Mechanical Engineering, University of Bristol, BS8 1FD, United Kingdom}
\author{John~G.~Rarity}
\affiliation{Quantum Engineering Technology Labs, School of Electrical, Electronic and Mechanical Engineering, University of Bristol, BS8 1FD, United Kingdom}

\date{\today}
\begin{abstract}

Entanglement, a key feature of quantum mechanics, is recognized for its non-classical correlations which have been shown to provide significant noise resistance in single-photon rangefinding and communications. Drawing inspiration from the advantage given by energy-time entanglement, we developed an energy-time correlated source based on a classical laser that preserves the substantial noise reduction typical of quantum illumination while surpassing the quantum brightness limitation by over six orders of magnitude, making it highly suitable for practical remote sensing applications. A frequency-agile pseudo-random source is realized through fibre chromatic dispersion and pulse carving using an electro-optic intensity modulator. Operating at a faint transmission power of 48~$\mu$W, the distance between two buildings 154.8182~m apart can be measured with a precision better than 0.1~mm,
 under varying solar background levels and weather conditions with an integration time of only 100~ms. These trials verified the predicted noise reduction of this system, demonstrating advantages over quantum illumination-based rangefinding and highlighting its potential for practical remote sensing applications.

\end{abstract}


\maketitle

\section{Introduction}
\noindent
Quantum entanglement has been widely explored for its potential to improve performance in the fields of communication~\cite{Chen2021Nature,Hu2021PRL}, computing~\cite{Graham2022Nature,Ladd2010Nature}, sensing~\cite{Guo2020NP,Hao2022PRL,Malia2022Nature,Xia2023NP,Zhang2021QST} and imaging~\cite{Defienne2021NP,Gregory2020SA}. In the field of quantum metrology, the method offers advantages including enhanced precision~\cite{Giovannetti2004Science,Toth2012PRA,Triggiani2023PRA} and noise suppression~\cite{Liu2019Optica,Lyons2018SA,Nair2011PRL,Zhang2015PRL,Zhuang2023NP}, which are particularly critical in remote sensing under challenging conditions, such as strong background and increased transmission loss. Since Lloyd’s pioneering work in 2008~\cite{Lloyd2008Science}, quantum illumination has garnered significant attention for its superior noise reduction capabilities in noisy and lossy environments. This improvement is primarily attributed to the entanglement and correlation properties of photon pairs~\cite{Shapiro2009njp,Tan2008PRL}. Lloyd’s theory projected that using $n$-dimensional entangled photon states could reduce noise compared to unentangled single-photon states. Subsequent studies~\cite{Zhang2015PRL,Guha2009PRA, Frick2020OE} have explored splitting quantum sources into $n$ frequency modes, allowing quantum systems to filter out unentangled noise via energy-time entanglement measured using joint detection, and demonstrating an enhanced signal-to-noise ratio (SNR).

However, applying quantum illumination to practical target detection presents significant challenges~\cite{Zhuang2017PRA,Zhuang2017PRL}. Generating, maintaining and detecting entangled photon pairs are more complex processes than their classical analogues, making quantum setups less robust. For instance, entanglement-based detection can require quantum memory to store the reference photon, complicating the detection platform and requiring prior knowledge of the target’s distance, which limits its applicability in rangefinding. The target in any rangefinding demonstration is a classical object that is either there or not, which in the language of quantum information means it is in the computational basis. As such, correlation measurements (without entanglement) can retain the key noise-reduction benefits while simplifying the system~\cite{Chang2019APL,Frick2020OE,Lopaeva2013PRL}. These advancements eliminate the need for pre-existing range information and quantum memory, making the setup more practical for distance measurement. Typical setups generate correlated photon pairs using a nonlinear crystal \cite{Zhang2015PRL,Frick2020OE,Rarity1990AO,Zhang2020PRA}  or semiconductor optical chip~\cite{Liu2019Optica}, sending one photon of each pair (signal) to the target while keeping the other (idler) as a reference. Range can then be measured using time differences between the two photons of each pair. Uncorrelated noise is filtered out via energy-time correlations, noticeably boosting detection sensitivity.

\begin{figure*}
    \includegraphics[width=2.05\columnwidth]{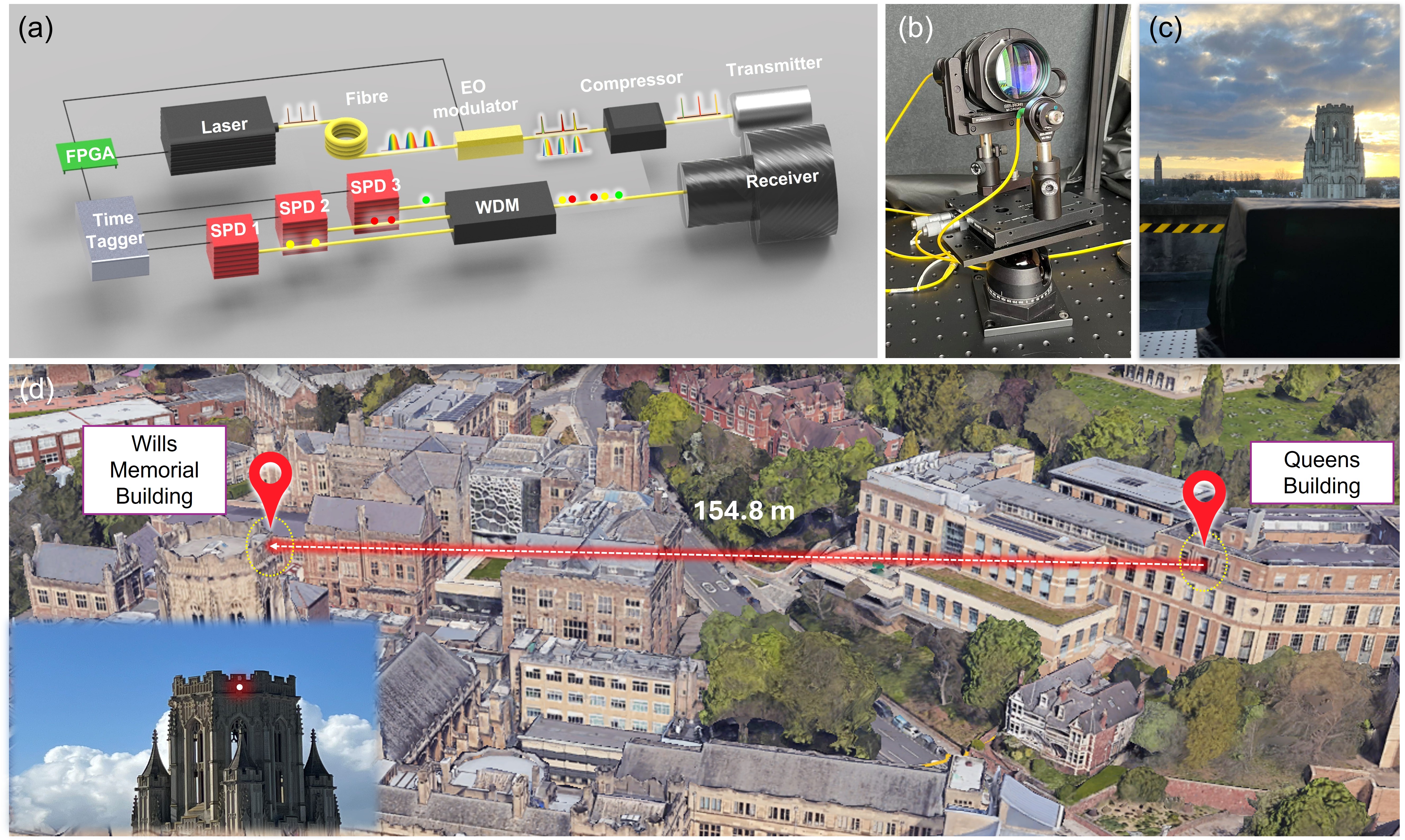}
    \caption{\textbf{Experimental setup. $|~$}(a)~Experimental setup for the rangefinding field trial. (b) The front view shows the transmitter and receiver on the balcony, connected to the lab by two single-mode fibres. (c) The back view depicts the black box housing the transmitter and receiver aimed at the target (the external wall of the Wills Memorial Building (WMB)). (d) A 3D Google map shows the distance between the balcony (Queen's Building) and the target (WMB), with the bright white spot in the inset indicating the illumination region of the target. Underlying map from Google Earth.}
    \label{fig:Fig/Fig2}
\end{figure*}

Despite these improvements, the brightness of photon-pair sources remains a significant limitation. Brightness is critical for remote target detection, as the number of detected back-scattered photons diminishes quadratically with distance, and at longer ranges, this can extend to a quartic relationship due to light divergence. The brightness of commonly used spontaneous parametric down-conversion (SPDC) sources is inherently limited by multi-photon emission, which when avoided results in low emission rates. Semiconductor quantum dot sources offer more efficient photon generation, but extracting photons is still challenging due to the high refractive index of surrounding materials~\cite{Nie2021APL}. Additionally, the maximum detection rate is constrained by the number of photons per time resolution window (typically $>$100~ps). For example, with correlated photon pair emission into $n$ modes and detection using $n$ single-photon detectors, the maximum count rate can reach $10^{10}$ photons per second (assuming $n=10$ and 0.1 pairs per detection window). This limits the maximum system loss to $\sim80$~dB, considering an overall system efficiency of $-20$~dB, which restricts the practical use of quantum illumination in remote sensing. In contrast, classical sources can easily achieve many orders of magnitude higher brightness levels~\cite{Buller2005RSI,Xu2021Optica}, making them more suitable for long-range detection. For example a 600 mW average power laser with a repetition rate of 500 kHz results in $10^{13}$ photons per pulse, has been reported for imaging over 200~km~\cite{Xu2021Optica}.

Recent research has explored the idea of using energy-time correlations from coherent sources to maintain quantum-like advantages~\cite{Kaltenbaek2008NP,Liu2023NC,Qian2023PRL}. However, these systems rely on nonlinear properties, which introduce a similar complexity to quantum sources and confine the detection range to be within the coherence length. As a result, all of them are demonstrated under controlled laboratory conditions and are difficult to extend to real-world remote sensing.

To address these limitations, we have developed a robust system that generates energy-time correlations without requiring a frequency conversion process, by modulating a classical source, while maintaining the noise reduction benefits of quantum illumination. This approach validates our theoretical model (Eq.~S10 in Supplementary Section II) where we find that the SNR simplifies to

\begin{equation}
    SNR  =\frac{\mathcal{S}}{\sqrt{\mathcal{S} + \mathcal{D}/n + \mathcal{B}/n}} \, .
    \label{equ_SNR_n_channel}
\end{equation}

\noindent where $\mathcal{S}$, $\mathcal{D}$ and $\mathcal{B}$ are the total correlated signal counts, dark counts and background counts across all channels. We clearly see that the dark and background counts are suppressed by using multiple channels $n$, resulting in an enhanced SNR (see Supplementary Section II for full derivation).
Using a three-channel ($n = 3$) frequency-agile correlated source based on fibre chromatic dispersion, we demonstrate noise reduction through field trials under various weather conditions over distances of more than 150~m. With the transmitter power attenuated to less than 50~$\mu$W the system shows potential for eliminating crosstalk in multi-use conditions, or for covert illumination, while demonstrating significantly higher brightness than entangled photon sources. Notably, the source power can be easily increased, making it suitable for longer-range measurements and showing that the brightness limitations of quantum systems can be bypassed without sacrificing noise suppression.

\section{Results}

\begin{figure}
    \centering
    \includegraphics[width=1\columnwidth]{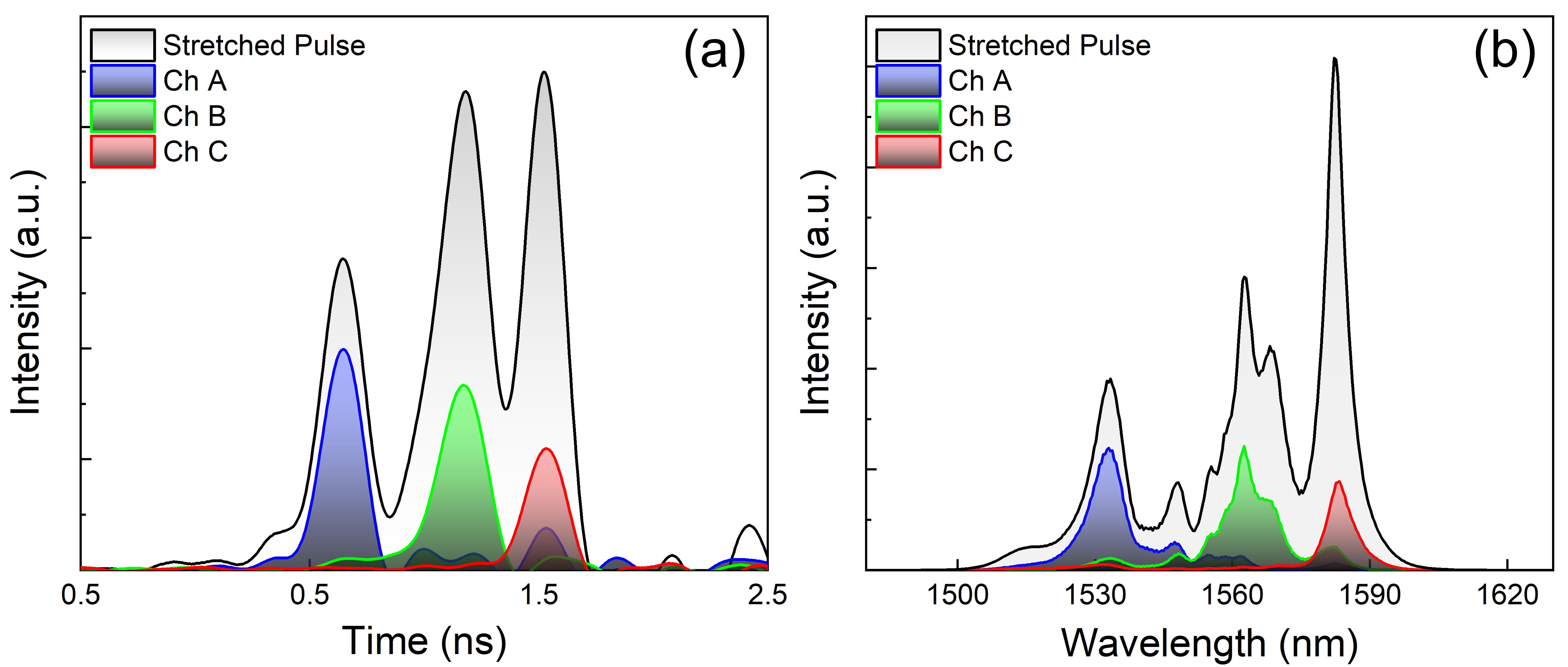}
    \caption{\textbf{Energy-time source characterisation. $|~$} (a)~Time-domain pulse traces showing the separation of three selected channels (coloured regions) from the stretched pulse (grey area with black line) originating from the initial fs-laser pulse. (b) Spectral separation of the three energy channels (coloured regions) derived from the original classical light source (grey area with black line).}
    \label{fig:Fig/Fig3}
\end{figure} 

Energy-time correlations in our quantum-inspired source are built up by stretching femtosecond laser pulses to nanosecond scale using fibre chromatic dispersion then pulse carving to create three frequency channels as described in the Online Methods section and illustrated in Fig.~\ref{fig:Fig/Fig2}. Figures~\ref{fig:Fig/Fig3}(a) and (b) display the characterization of the created correlation with respect to time and energy, respectively. The grey area with a black line represents the stretched pulse of the primary fs-laser before temporal modulation, while the coloured areas correspond to the three selected channels after electro-optic intensity modulator (EOIM) modulation. 
Figure~\ref{fig:Fig/Fig3}(a) demonstrates the time-domain selection and separation of the three channels by applying different time delay windows to the stretched pulse.
Figure~\ref{fig:Fig/Fig3}(b) shows the energy separation of the three channels, characterized by their spectra with central wavelengths of 1532.2~nm, 1562.5~nm, and 1583.7~nm.
The direct correspondence between the pulse time sequence and the increasing wavelength spectra demonstrates the successful creation of energy-time correlation across the three channels. The three wavelengths are selected using a pseudorandom sequence of length 10~$\mu$s to allow unaliased rangefinding up to 1.5~km distance. After successfully creating three separate wavelengths by time-gating, the short pulse is reformed in a grating-based compressor.

The quantum-inspired correlation source achieves an average power of 147.5~$\mu$W before the fibre splitter, corresponding to 23~million photons per pulse. This represents more than 60~dB improvement compared to conventional quantum sources that are forced to operate at one photon per time window on average~\cite{Frick2020OE}. 
This significant enhancement in source brightness has enabled us to extend the detection distance from several metres in the lab~\cite{Frick2020OE} to field trials conducted between two buildings, separated by a distance of over 150~metres.

\begin{figure}
    \centering
    \includegraphics[width=1\columnwidth]{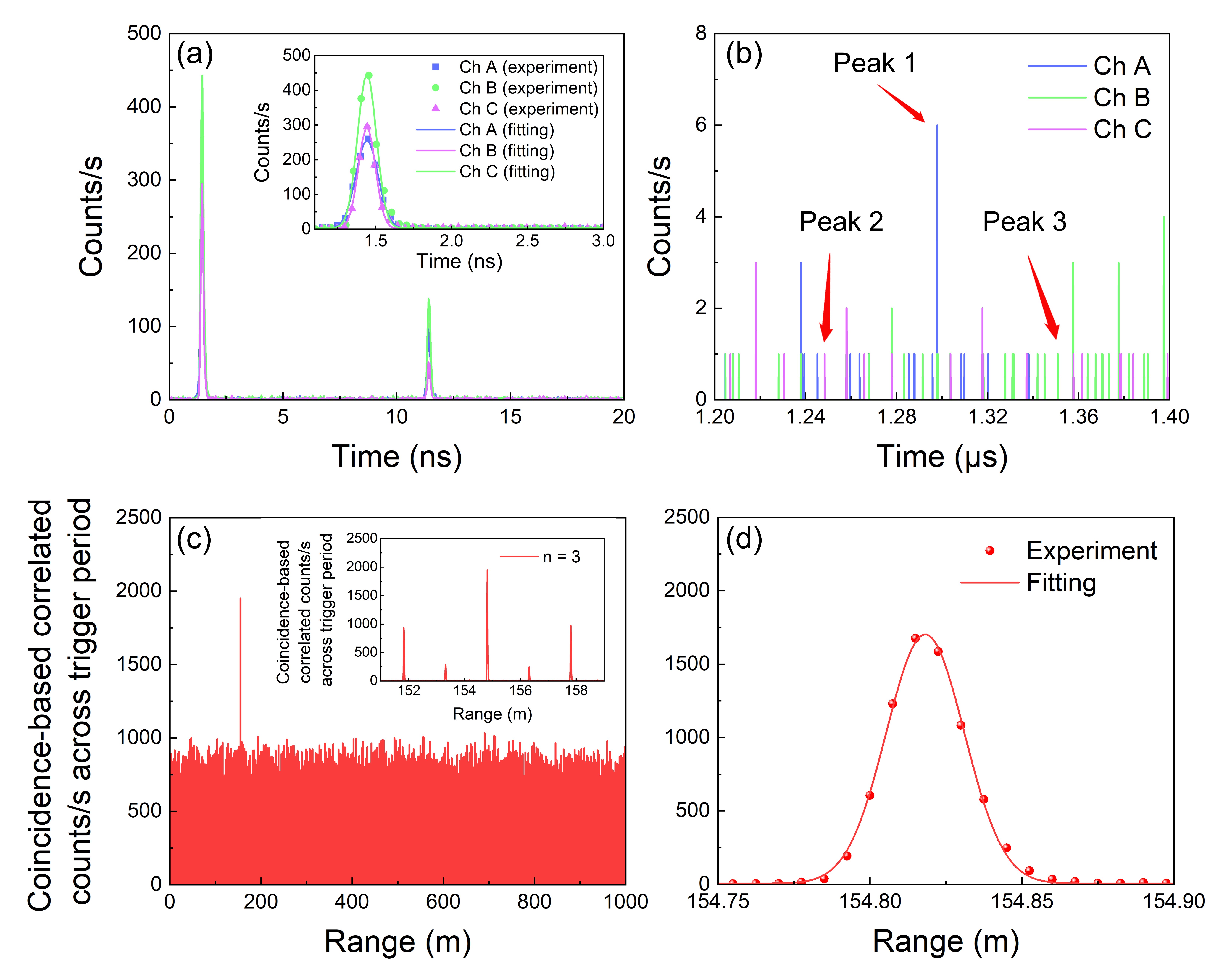}
    \caption{\textbf{Nighttime characterisation results. $|~$} (a)~Return photon-counting histograms for each channel (A, B, and C) triggered at the source repetition rate of 50~MHz, displayed within one source period of 20~ns with one second integration time. Inset: magnified view of the primary signal peak. (b) A 200~ns excerpt of a return photon-counting histogram triggered at the pattern repetition rate of 100~kHz, integrated over one second. Peaks 1, 2 and 3 correspond to echo signal photons, modulator leaky photons, and dark counts, respectively. (c) Coincidence-based cross-correlation between the normalized reference electrical signal and the detected echo photons back from WMB, with an integration time of one second. The highest peak at 154.81822(5)~m marks the measured range. Inset: expanded view of the main peak, showing periodic pulses from the original source. (d) The main peak, further magnified from the inset of (c), fitted with a Gaussian function (red line).}
    \label{fig:Fig/Fig4}
\end{figure} 

The results from night-time measurements, where photons were scattered back from the external wall of Wills Memorial building (WMB) with an integration time of one second, are shown in Fig.~\ref{fig:Fig/Fig4}. Figure~\ref{fig:Fig/Fig4}(a) presents the photon-counting histograms of return photons in three channels within one source period of 20~ns. The leftmost peaks of the three channels are the main return signals and are well aligned with each other, indicating that the energy-time correlation has been effectively concealed by the grating-based compressor.
The right peaks are the breakthrough from the EOIM caused by reducing the repetition rate from 100~MHz to 50~MHz. This leakage arises from the relatively low EOIM extinction ratio, which is due to the broadband spectra in each channel. This issue could be mitigated by increasing the number of channels, thereby narrowing the spectral width of each channel. Figure~\ref{fig:Fig/Fig4}(b) shows a~200 ns excerpt of a 10~$\mu$s long photon-counting histogram, triggered by the start of the random pattern. Transmission losses scattered from the Lambertian target were calculated to be 98.1~dB in this experiment, resulting in a maximum count rate of only 6 photons per time bin, some of which may include background noise or detector dark counts. The smaller peaks with counts around one photon are due to leakage from the EOIM, dark counts, and background noise. By post-selecting the encoded timing information, we can identify that peak~1 corresponds to the echo signal, peak 2~is EOIM leakage, and peak~3 is detector dark counts. Given the night-time conditions, the background noise level was low and can be considered negligible.

Distance information was extracted from the photon-counting histograms of the return signal using the coincidence-based correlation method.
The coincidence counts in each channel were calculated by cross-correlating the normalized reference, made up of the random channel sequence repeating for the chosen integration time, with the echo photon counts for each channel, resulting in three-channel coincidence counts shown in Fig.~\ref{fig:Fig/Fig4}(c). From this data the round-trip time was estimated from the location of the highest peak, and the corresponding range was calculated using the formula $r = ct/2$, where $c$ is the speed of light and $t$ is the round-trip time of the illuminating photons. Magnifying the peak region in the inset, several periodic side peaks can be observed. The relatively high side peaks are due to the 50~MHz pulse repetition period, while the smaller peaks arise from leakage of the 100~MHz pulse. The measured peak (with a time bin of 50~ps) was fitted using a Gaussian function (see Fig.~\ref{fig:Fig/Fig4}(d)), centred at a distance of 154.81822 $\pm$ $4.8\times10^{-5}$ m. This fitting improves precision to $\pm~48~ \mu$m, overcoming the limitations set by detector jitter. 
In contrast the single shot depth resolution is determined by the histogram full width at half maximum (FWHM) of the sum of the three-channel pulses shown in Fig.~\ref{fig:Fig/Fig4}(d), which is 25.69 $\pm$ 0.097~mm. 
Additionally, the unambiguous range was extended from 3~m to 1,500~m, set by the length of the random channel sequence. This breaks the limit of a periodic source on maximum detection distance, caused by the high repetition rate of 50~MHz, and could be further extended using a real-random sequence. 

\begin{figure}
    \centering
    \includegraphics[width=1.0\columnwidth]{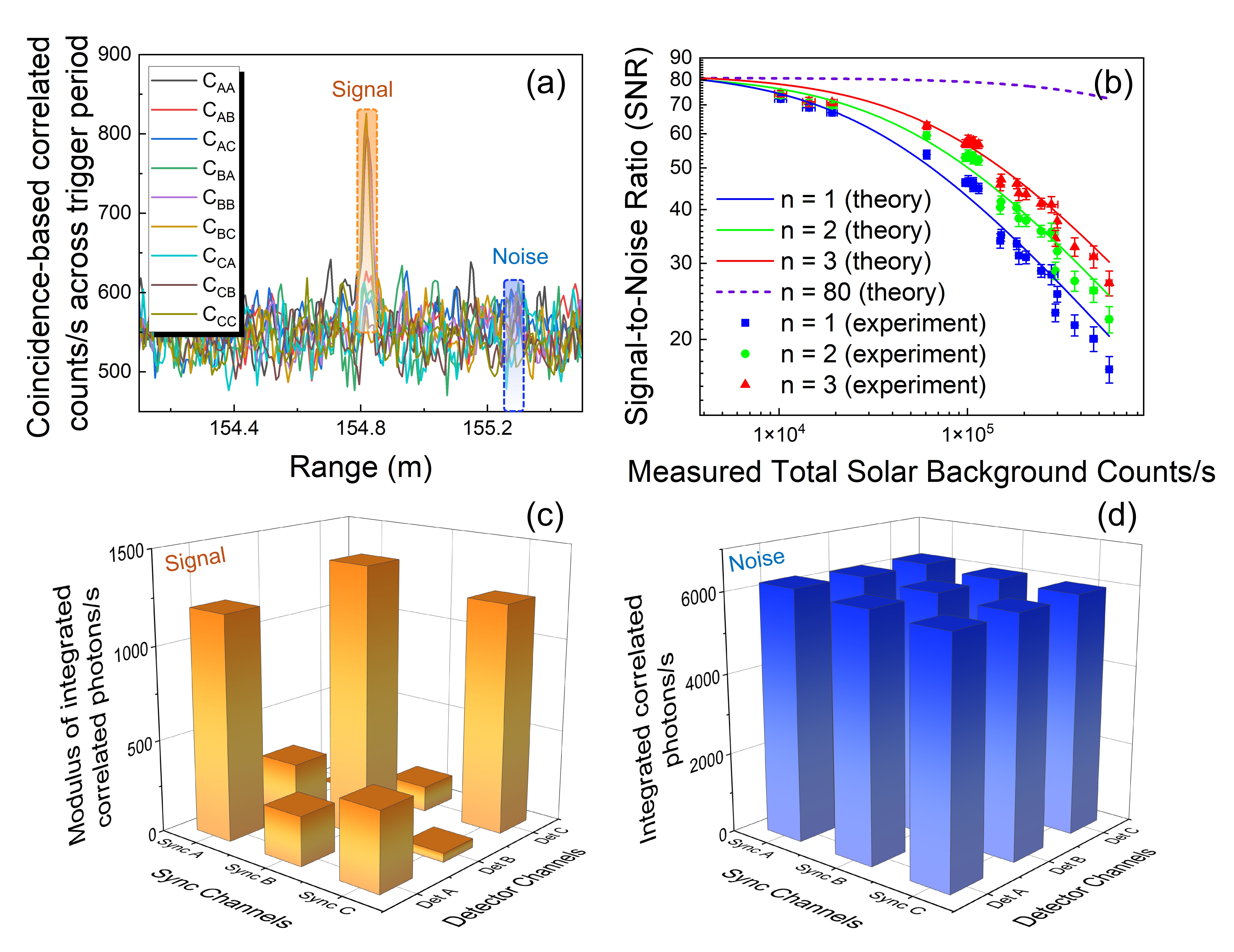}
    \caption{\textbf{Daytime enhancement to SNR. $|~$} (a)~Coincidence-based cross-correlation peaks $C_{ij}$ between the normalized reference channels $i$ and detected photon channels $j$ ($i, j \in A, B, C$) collected in bright daylight (sunshine falling on WMB) and one second integration time. (b) Signal-to-noise ratio for one-channel ($n = 1$), two-channel ($n = 2$), and three-channel ($n = 3$) platforms, integrated over one second, as a function of the detected total solar background. Symbols represent the experimental data, with the fitted theoretical model shown as coloured curves. The purple dashed line shows the simulated results for 80 channels using commercial dense wavelength division multiplexing (DWDM). (c,d) Correlation tomography extracted from the orange signal region and blue noise region in (a), respectively.}
    \label{fig:Fig/Fig5}
\end{figure} 

To experimentally validate the theoretical model that shows a significant noise reduction effect from using multiple energy channels, we compared a one-channel ($n = 1$), two-channel ($n = 2$) and three-channel ($n = 3$) system under varied solar background levels using data from three single-photon detectors. Since the periodic peaks are primarily due to the source repetition rate, we focused on the range peak region within a 1.5 m (10 ns) window, excluding the influence of the periodic signal.

\begin{figure*}
    \includegraphics[width=2.05\columnwidth]{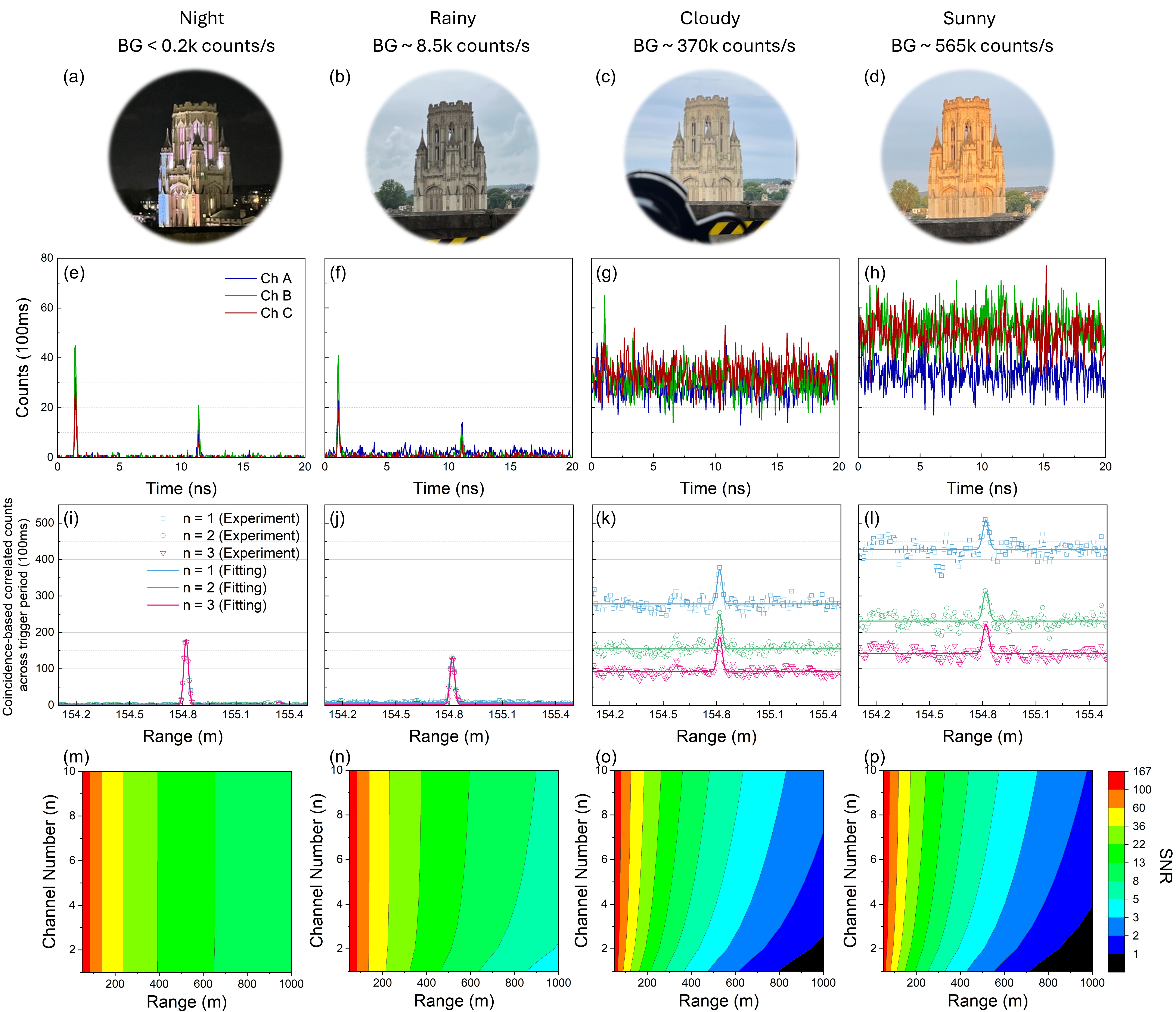}
    \caption{\textbf{Field trial rangefinding results. $|~$} (a-d)~Images of the target building (WMB) under various weather conditions (BG: Total background counts measured by three single-photon detectors). (e-h) Echo photon-counting histograms for each channel (A, B, and C), displayed within single source period of 20~ns for the corresponding weather scenarios, with an integration time of 100~ms. (i-l) Coincidence-based cross-correlated peaks under different weather conditions using one channel ($n = 1$), two channels ($n = 2$), and three channels ($n = 3$), integrated over 100~ms. (m-p) Simulation results of SNR under corresponding background level as a function of channel number ($n$) and the detection range. 
    These results demonstrate that significant noise reduction can be achieved by increasing channel numbers, particularly in strong daytime solar background.}
    \label{fig:Fig/Fig6} 
\end{figure*}

First, coincidence-based correlated counts $C_{ij}$ were calculated between the normalized reference channel $i$ and detected photon channel $j$ ($i,j \in {A,B,C}$) using data collected during the daytime with 1~s integration time, as displayed in Fig.~\ref{fig:Fig/Fig5}(a). Here, correlated elements occur when $i=j$, and uncorrelated elements occur when $i\neq j$. For single-channel measurements ($n=1$), no energy-time correlation exists, meaning all elements are considered:
\begin{equation}
\begin{split}
    C_{total,n=1} & = C_{AA} + C_{AB} + C_{AC} \\
                 &   + C_{BA} + C_{BB} + C_{BC} \\
                 &   + C_{CA} + C_{CB} + C_{CC}.
    \label{n=1}
\end{split}
\end{equation}

For two-energy channels ($n=2$), two of the three photon detection channels (e.g., A and C) are treated as one combined channel. The correlated counts between this combined channel and the third channel (e.g., B) are calculated as follows:
\begin{equation}
\begin{split}
    C_{total,n=2} & = (C_{AA} + C_{AC} + C_{CA} + C_{CC})  + C_{BB}\\
\end{split}
\label{n=2}
\end{equation}

Finally, for three-energy channels ($n=3$), the correlated counts for all three channels are given by:
\begin{equation}
\begin{split}
    C_{total,n=3}  = C_{AA} + C_{BB} + C_{CC}.
\end{split}
\label{n=3}
\end{equation}

As illustrated in the correlation tomography of Figs.~\ref{fig:Fig/Fig5}(c) and (d), the signal and noise counts were extracted from Fig.~\ref{fig:Fig/Fig5}(a) by integrating the peak region, with the average noise level subtracted (orange area) and the fluctuating region, estimated as the average noise level multiplied by the integration time (blue area). In Fig.~\ref{fig:Fig/Fig5}(c), the signal counts primarily come from correlated elements ($i=j$), indicating low crosstalk across multiple energy channels. Uncorrelated counts ($i\neq j$), which are primarily induced by time jitter, can be minimized by increasing the delay between adjacent channels. 
Notably, the signal displayed in the correlation tomography (Fig.~\ref{fig:Fig/Fig5}(c)) has been background-subtracted to reflect the net signal used in the SNR calculation Eq.~\eqref{equ_SNR_n_channel}. As shown in Eqs.~\eqref{n=1},~\eqref{n=2}, and~\eqref{n=3}, the signal remains nearly constant across different configurations due to low crosstalk. In contrast, the noise counts depicted in Fig.~\ref{fig:Fig/Fig5}(d) are comparable between the correlated and uncorrelated elements, indicating that the total noise decreases approximately proportionally to the number of channels ($n$), owing to the reduced number of contributing elements in the noise term.

Based on the above processing, the SNR of rangefinding systems with varying channel numbers ($n \in {1,2,3}$) was measured with one-second integration under different levels of solar background during daylight, as demonstrated in Fig.~\ref{fig:Fig/Fig5}(b). The experimental data, marked as symbols with different colours, were fitted to our theoretical model (Supplementary Section II), with coloured curves representing different channel numbers ($n$). The agreement between the model and field trial results demonstrates the validity of the theoretical framework. The SNR improves with an increasing number of channels, particularly under higher background rates. A more pronounced noise reduction can be realized by extending the channel quantity to 80 using off-the-shelf dense wavelength division multiplexing (DWDM) components, as indicated by the purple theoretical dashed curve, with an enhancement of 5.7 dB at the maximum detected solar background level of 565 kcounts/s. 

Experimental results under various weather conditions conducted on the same day are presented in Fig.~\ref{fig:Fig/Fig6}. Figures~\ref{fig:Fig/Fig6}(a)-(d) display photos of the WMB taken from Queens Building Balcony (QBB), along with the corresponding weather conditions and detected total background counts from three single-mode fibre coupled single-photon detectors. Figures~\ref{fig:Fig/Fig6}(e)-(h) show the return photon histograms for each channel, integrated over 100~ms and displayed within a period of 20~ns. Using the previously described coincidence-based cross-correlation method, the range peaks can be identified in Figs.~\ref{fig:Fig/Fig6}(i)-(l). During nighttime, the echo photon peak is notably clear, with an extremely low noise level in Fig.~\ref{fig:Fig/Fig6}(e), corresponding to a distinct range peak centred at 154.81822(5)~m (Fig.~\ref{fig:Fig/Fig6}(i)). Although the building appears colourful at night in Fig.~\ref{fig:Fig/Fig6}(a) due to visible light illumination, its influence on noise counts is negligible, as the visible light is filtered out by a long-pass filter and cannot be detected by the InGaAs single-photon detectors. During rainy moments, the detected background level increased to 8.5~kcounts/s, resulting in a dark appearance of the building in Fig.~\ref{fig:Fig/Fig6}(b). Consequently, the noise level in the photon counting histogram (Fig.~\ref{fig:Fig/Fig6}(f)) rises, while the signal peak slightly decreases due to higher transmission loss from the rain drops. As shown in Fig.~\ref{fig:Fig/Fig6}(j), the influence on the range peak becomes apparent, evidenced by the reduction of peak intensity. Under cloudy conditions, with measured background counts reaching 370~kcounts/s, the building appears brighter compared to the rainy period, as depicted in Fig.~\ref{fig:Fig/Fig6}(c). The background intensity in both photon-counting histogram (Fig.~\ref{fig:Fig/Fig6}(g)) and correlated photon histogram (Fig.~\ref{fig:Fig/Fig6}(k)) increases. However, the signal peak (after subtracting the noise level) is diminished due to the elevated background level. When sunlight illuminates the building at sunrise, as shown in Fig.~\ref{fig:Fig/Fig6}(d), the three signal peaks in the photon-counting histogram (Fig.~\ref{fig:Fig/Fig6}(h)) become indistinguishable due to strong solar background noise. Nevertheless, the range peak remains detectable with the three-channel platform, while the single-channel peak becomes challenging to resolve (Fig.~\ref{fig:Fig/Fig6}(l)). By comparing results from different numbers of energy channels, significant noise suppression is observed with the three-channel peak ($n = 3$), particularly under sunny conditions with higher background counts (Fig.~\ref{fig:Fig/Fig6}(l)). The corresponding simulation results using the measured solar background level are shown in Figs.~\ref{fig:Fig/Fig6}(m)-(p). The SNR is increased significantly with longer distance measurements due to increased transmission loss, especially under strong background levels. 
It should be noted that the current system can be further optimized through reductions in transmission losses and increased source power, potentially extending the detection range beyond 1~km.

\section{Discussion}

In this work, we demonstrated the generation of energy-time correlations using a modified classical source, inspired by quantum illumination principles for noise reduction. By utilizing fibre chromatic dispersion and time-delay selection via an EOIM, multiple energy channels with corresponding time delays were generated. Consequently, energy-time correlations were successfully verified in a classical source using three isolated channels. The brightness of this developed source shows a remarkable improvement by a factor of $10^6$ compared to quantum sources, effectively overcoming the distance limitations typically associated with such systems, and can be further enhanced easily. For instance, maximizing the extinction ratio of the electro-optic intensity modulator (EOIM) by increasing the electrical signal amplitude, combined with the integration of an optical amplifier, could provide a gain of up to 23 dB. System losses can also be reduced by employing multi-mode fibre collection in conjunction with a larger-aperture telescope. The brightness enhancement extends the detection range to a Lambertian target (the external wall of the WMB), from several metres in a laboratory setting to over 100 metres in outdoor environments. Our theoretical model indicates that the noise reduction benefit seen in quantum rangefinding can be retained - and even enhanced - by increasing the number of channels, as the dark count noise remains unaffected by the addition of extra detectors. This overcomes the limitation of quantum rangefinding, which is constrained by an optimal number of channels. Field trials conducted under various daytime weather conditions including bright sunlight confirmed significant SNR improvement with the three-channel configuration, particularly in high solar background conditions, with a measured range peak at 154.81822(5)~metres.

\begin{figure}
    \centering
    \includegraphics[width=1\columnwidth]{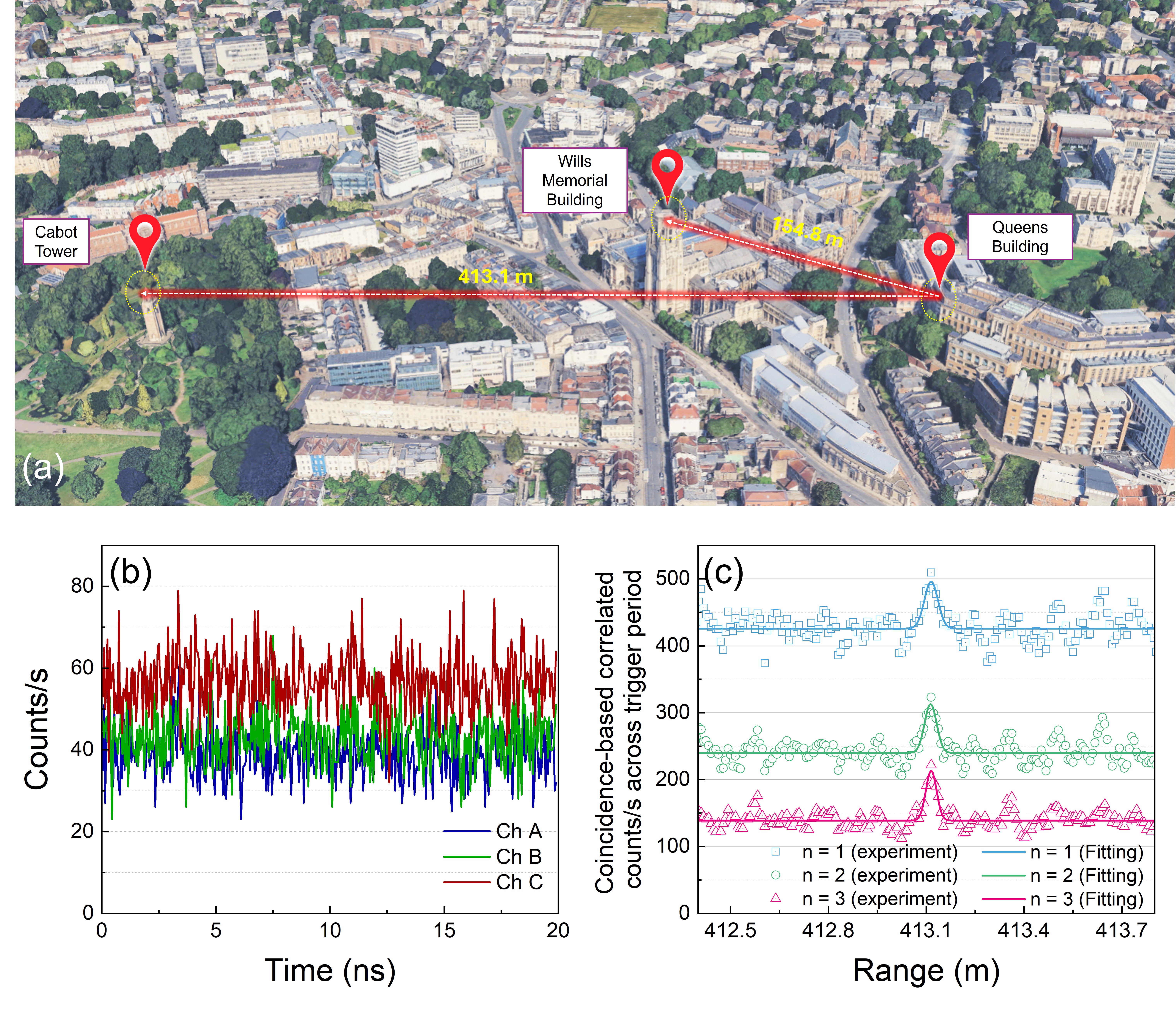}
    \caption{\textbf{Extended field trial rangefinding to over 400~m. $|$} (a) A 3D Google map shows the distances between the balcony (Queen’s Building) and the targets (both Wills Memorial Building and Cabot Tower) (b) Return photon-counting histograms for each channel (A, B, and C) from Cabot Tower triggered at the source repetition rate of 50~MHz, displayed within one source period of 20~ns with one second integration time. (c) Coincidence-based cross-correlated peaks using n channels ($n$ = 1, 2 and 3, respectively), with an integration time of one second. The highest peak at 413.1151~m marks the measured range, fitted with a Gaussian function. Underlying map in (a) from Google Earth.}
    \label{fig:Fig/Fig7}
\end{figure} 

In our field trials, the detection range was constrained by the available experimental space rather than the brightness of the source, indicating that further extension is possible. This potential has been demonstrated in subsequent ranging and imaging experiments, with preliminary results shown in Fig.~\ref{fig:Fig/Fig7}. The new target is located 413.1~metres away from our transmitter, as shown in the 3D Google Map in Fig.~\ref{fig:Fig/Fig7}(a)). The distance was measured using the same commercial rangefinder used for measuring the distance to WMB with the resolution of 0.1~m. The return photon histograms for each channel (integrated over one second and displayed within a 20~ns window in Fig.~\ref{fig:Fig/Fig7}(b)) and the correlated range peaks using multiple channels ($n$ = 1, 2, and 3 in (c)) were detected at a distance of 413.1151~metres under bright solar background conditions, demonstrating the system’s potential for extending the detection range during daytime. Consistent with previous results in Figs.~\ref{fig:Fig/Fig6}(h) and (l), the range peak using three channels remains clearly distinguishable, even when the single-channel signal is obscured by solar background noise. Further improvements can be achieved by the incorporation of dense wavelength-division multiplexing (DWDM) components with 80 channels which could significantly improve the SNR, offering a promising avenue for enhancing system performance under challenging environmental conditions.

The developed source also holds potential for covert rangefinding, leveraging another key advantage of quantum illumination. For instance, the transmission power is set to a low level of 48~$\mu$W, with pulse compression applied to obscure the generated correlations. To achieve fully covert operation, the unique signature of the laser source requires further manipulation. For example, the detectable repetition rate can be randomized by introducing random time delays, and the laser source can be replaced with a triggered thermal source to replicate the super-Poissonian statistics of solar background. These modifications would significantly enhance covertness and reduce the likelihood of detection. Similarly the low powers needed coupled with pseudo random coding could provide immunity to cross-talk in automotive Lidar applications.


\section{Online Methods}

\subsection{Entanglement-inspired source}

A schematic of our system is depicted in Fig.~\ref{fig:Fig/Fig2}(a). We used a fibre laser to create 58~fs long pulses at a repetition rate of 100~MHz with a central wavelength of 1563~nm and a spectral width of 67~nm.  The power was carefully controlled using a fibre-coupled digital variable attenuator. The short pulses were temporally stretched to 1.2~ns using a 1~km long telecommunication band single-mode fibre (SMF28). After passing through a fibre polarization controller, the pulse was directed to an EOIM for energy channel selection. Based on the fibre chromatic dispersion, three wavelength channels were picked up within 250~ps time windows distributed over the 1.2~ns pulse duration, allowing for precise channel selection via different time delays processed by a field-programmable gate array (FPGA). By encoding a pseudo-random pattern to the time delay in the electrical signal, wavelength channels were randomly selected. Using this method, energy-time correlations were successfully generated from a classical source. The pulse was then split by a 1x2 fibre coupler with a coupling ratio of 50:50, with one path directed to a grating-based compressor used to erase the channel-selection information for covert sensing and improving distance measurement resolution. The other path was either detected by a  fibre-coupled InGaAs photodiode (Thorlabs DET08CFC) connected to a fast oscilloscope (KEYSIGHT DSO-S 404A) for real-time performance monitoring, or sent directly to an optical spectrum analyser (OSA) for spectral analysis.

The synchronized electrical signal fed to the EOIM was initially derived from the fs-laser (operating at 100~MHz with a pulse width of 370~ps) and broadened to 4~ns using a pulse shaper. This broadened signal was then processed by a Xilinx ZYNQ-7000 FPGA, which it down-converted to a 50~MHz clock signal. The FPGA encoded a pseudo-random pattern of varying time delays via its internal IDELAY module onto 500 consecutive pulses, leading to a pattern that repeated every 10~$\mu$s. Subsequently, the electrical pulses were compressed to 250~ps and amplified to 3~V before being sent to the EOIM, which functioned as a high-speed optical switch, randomly selecting wavelength channels based on the encoded pattern. Another down-converted signal with a 100~kHz clock, corresponding to the pattern repetition rate, was sent to the time tagger (HydraHarp 400) as a reference.

\subsection{Field trail of entanglement-inspired rangefinding}

By integrating the developed source into a rangefinding platform, the signal was transmitted through a single-mode fibre to a collimator positioned on the QBB, as depicted in Fig.~\ref{fig:Fig/Fig2}(b). The average transmitter power was attenuated to 48~$\mu$W for the field trial. The system had a full beam divergence of 0.032$^{\circ}$ (0.56~mrad), and the signal illuminated the external wall of WMB, approximately 154.8~m away from QBB, as shown in Fig.~\ref{fig:Fig/Fig2}(c). The measured distance, shown in Fig.~\ref{fig:Fig/Fig2}(d), was determined using a commercial rangefinder with a resolution of 0.1~m. The backscattered light from the building was collected by a 48~mm telescope and coupled into a single-mode fibre, passing through a long-pass filter at 1500~nm to block visible and near-infrared solar background. This filtered signal was split by a wavelength division multiplexer (WDM) and directed into three infrared single-photon detectors (ID230) through single-mode fibres, as illustrated in Fig.~\ref{fig:Fig/Fig2}(a). These InGaAs/InP photon detectors, set with a 2~$\mu$s dead time and $15\%$ quantum efficiency, sent their detected signals to a Time Tagger (Picoquant HydraHarp 400), synchronized with the beginning of the repeating pattern down-converted by the FPGA. The field trial ran continuously from night through to day under varying weather conditions, enabling us to perform a comparison of the experimental results with our noise-reduction theory.

\begin{acknowledgements}
The authors acknowledge support from EPSRC through the QuantIC (EP/T00097X/1), the Quantum Position, Navigation and Timing (QEPNT) Hub (EP/Z533178/1), and the Quantum Sensing, Imaging and Timing (QuSIT) Hub (EP/Z533166/1). ASC acknowledges support from The Royal Society (URF/R/221019, RF/ERE/210098, RF/ERE/221060).
\end{acknowledgements}

\bibliography{mylibrary}

\end{document}



\title{Supplementary information: Entanglement-inspired frequency-agile rangefinding}

\author{Weijie~Nie}
\email{weijie.nie@bristol.ac.uk}
\affiliation{Quantum Engineering Technology Labs, School of Electrical, Electronic and Mechanical Engineering, University of Bristol, BS8 1FD, United Kingdom}
\author{Peide~Zhang}
\affiliation{Quantum Engineering Technology Labs, School of Electrical, Electronic and Mechanical Engineering, University of Bristol, BS8 1FD, United Kingdom}
\author{Alex~McMillan}
\affiliation{Quantum Engineering Technology Labs, School of Electrical, Electronic and Mechanical Engineering, University of Bristol, BS8 1FD, United Kingdom}
\author{Alex~S.~Clark}
\affiliation{Quantum Engineering Technology Labs, School of Electrical, Electronic and Mechanical Engineering, University of Bristol, BS8 1FD, United Kingdom}
\author{John~G.~Rarity}
\affiliation{Quantum Engineering Technology Labs, School of Electrical, Electronic and Mechanical Engineering, University of Bristol, BS8 1FD, United Kingdom}

\date{\today}


\maketitle
\section{Energy-time correlations from pulse stretching}

The generation of energy–time correlations in the developed source is achieved through fibre-induced chromatic dispersion, followed by time-domain post-selection using an electro-optic intensity modulator (EOIM), as detailed in Methods Online. To elucidate the underlying mechanism, we begin by describing the temporal stretching of the pulsed laser in the dispersive fibre, which is governed by the wave-vector $k$,
as a function of the angular frequency $\omega$, expressed as follows
\begin{equation}
k(\omega)=k(\omega_0)+\beta_1\left(\omega-\omega_0\right)+\frac{1}{2} \beta_2\left(\omega-\omega_0\right)^2+\ldots
\label{equ_dispersion}
\end{equation}
where $\omega_0$ is the central angular frequency, while $\beta_1$ and $\beta_2$ describe the group delay and the quadratic group velocity dispersion, respectively. The transmission time $T$ for light with angular frequency $\omega$ travelling through a fibre of length $L$ is given by
\begin{equation}
T(\omega)=L\left|\beta_2\right| \omega-L\left|\beta_2\right|\omega_0+T\left(\omega_0\right)
\label{equ_correlation_T_omega}
\end{equation}
This expression can be reformulated in terms of photon energy $E$ to describe the energy-time correlation, where the transmission time $T$ as a function of photon energy $E$ is
\begin{equation}
T(E)=L\left|\beta_2\right| E-L\left|\beta_2\right|\frac{E_0}{\hbar}+T\left(\frac{E_0}{\hbar}\right)
\label{equ_correlation_T_E}
\end{equation}
where $E_0$ is the central photon energy, and $\hbar$ is the reduced Planck constant. By selecting different time delays within each pulse, different energy channels can be accessed.

\section{Signal-to-noise ratio enhancement}
\label{supp-SNR}

Building on the generated energy–time correlations in the broadband pulse, the pulses are selectively modulated at distinct time delays and encoded with a pseudo-random time sequence afterwards, enabling frequency agility in the developed source.
The resulting pulses in random wavelength channels are transmitted to a target, and reflected photons are collected, separated into different wavelength channels, and sent to single-photon detectors. We then calculate the time-delayed cross-correlation function to find the coincidence rate at different time delays $\tau$

\begin{equation}
   C(\tau) = \int_{0}^{T_t}\langle S(t) R(t+\tau) \rangle dt
   \label{equ_correlated_counts}
\end{equation}

where $S$ is the photon detection signal, $R$ is the normalized known reference signal from the energy-time correlation, and $T_t$ is the time over which the random pulses are sent. The histogram will show a peak corresponding to the time of flight of the reflected photon signal that can then be used to calculate the round-trip range to the target.


To evaluate the noise reduction capability of this entanglement-inspired $n$-channel source in a rangefinding system, we can calculate the signal-to-noise ratio (SNR) assuming the system is shot-noise limited as
\begin{equation}
    SNR = \frac{\sum_{i=1}^{n}C_{S_i}}{\sqrt{\sum_{i=1}^{n}C_{S_i}+\sum_{i=1}^{n}C_{N_i}}} \, ,
    \label{equ_SNR}
\end{equation}
\noindent where $C_{S_i}$ represents the signal photon coincidence counts integrated in a time window around the coincidence peak, and $C_{N_i}$ is the average coincidence noise counts during the same period. The return signal for each channel $i$ is
\begin{equation}
        C_{S_i} = \frac{N_S}{n} \eta_{i} \eta_S  \, ,
        \label{equ_signal}
\end{equation}
\noindent where $N_S$ is the total number of photons emitted by the transmitter spread across all $n$ channels, $\eta_{i}$ is the unbalance in channel power splitting with $\sum_{i=1}^{n}\eta_{i} = n$, and $\eta_S = \eta_t \eta_r \eta_d$ is the total channel efficiency taking into account the transmission and collection efficiency $\eta_t$, the reflection of the target $\eta_r$ and the detector efficiency $\eta_d$. We make the assumption here that these are the same for all channels as the difference between each channel is negligible. The primary noise in each channel is
\begin{equation}
    C_{N_i} = C_{B_i}+ C_{D_i} \, .
    \label{equ_noise}
\end{equation}
\noindent where $C_{B_i}$ is the background count from the environment, such as solar background, $C_{D_i}$ is the dark count from the single-photon detectors. The background count is
\begin{equation}
    C_{B_i}  = \frac{B_0 \cdot \Delta\lambda}{n} \frac{l \cdot T_p}{n} \eta_{B} = \frac{N_B}{n^2} \eta_{B} \, ,
    \label{equ_background}
\end{equation}
\noindent where $N_B$ is the total number of background photons present when the pulses are on and $\eta_{B} = \eta_c \eta_d$ is the lumped system efficiency for collecting ($\eta_c$) and detecting ($\eta_d$) background photons. We can see that $N_B$ is found by multiplying the background spectral density, $B_0$ in Hz/nm, and the total bandwidth of all energy channels, $\Delta \lambda$, normalised by the number of channels, $n$. This must then be multiplied by the total on-time of the reference pattern used in the cross-correlation that forms the coincidence histogram. The random pulse pattern has integer length $l$, pulses are shared between all channels $n$, and each pulse has a duration $T_p$. 
The dark counts from each photon detector are,
\begin{equation}
    C_{D_i} =\frac{D}{n}  \frac{l \cdot T_p}{n} = \frac{{N}_D}{n^2}\, ,
    \label{equ_darkcounts}
\end{equation}
\noindent where $D$ is the total dark counts of all single photon detectors, and the second term once again takes into account the time that the reference pattern pulses are on for each channel. Similar to background noise, We can write this as the all detectors dark counts when the total pattern is on, ${N}_D$, which is divided by a square of channel number $n$. 
Substituting Eq.~\ref{equ_background}, and Eq.~\ref{equ_darkcounts} into Eq.~\ref{equ_noise}, and substituting that and Eq.~\ref{equ_signal} into Eq.~\ref{equ_SNR} yields

\begin{figure}[t!]
    \centering
    \includegraphics[width=\columnwidth]{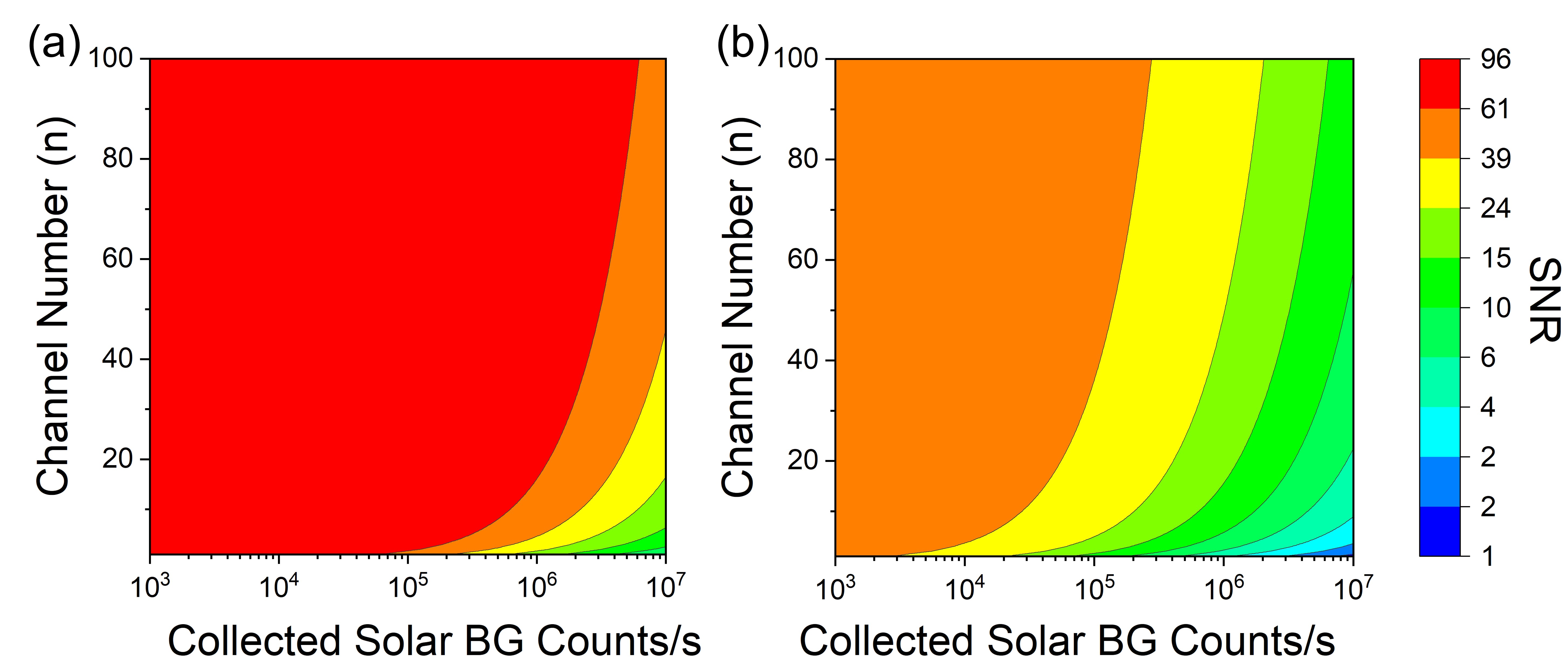}
    \caption{\textbf{Theoretical SNR improvements. $|$} Simulated SNR under high transmission losses of (a)~98.1~dB and (b)~104.8~dB, corresponding to a distance of 155~metres under sunny and rainy conditions, respectively, as a function of the number of channels and collected solar background (BG) counts per second. All parameters are set to the experimental values.}
    \label{fig:Fig/Fig1}
\end{figure}

\begin{align}
SNR &= \frac{N_S \eta_S}{\sqrt{N_S \eta_S + N_D/n + N_B \eta_B / n}} \notag \\
    &= \frac{\mathcal{S}}{\sqrt{\mathcal{S} + \mathcal{D}/n + \mathcal{B}/n}} \label{equ_SNR_n_channel}
\end{align}

\noindent where $\mathcal{S}$, $\mathcal{D}$ and $\mathcal{B}$ are the total correlated signal counts, dark counts and background counts across all channels (i.e., the counts recorded while the reference pattern is on). From this theoretical model, two distinct advantages of our approach over traditional quantum rangefinding~\cite{Frick2020OE} can be observed. First, the impact of background noise $N_B$ can be mitigated by increasing the number of channels $n$, allowing for significant noise reduction. Unlike quantum systems, where the optimal number of channels is constrained due to the contribution of dark counts in heralding detectors to the total noise, classical systems can exploit arbitrarily large numbers of channels without such limitations. Therefore, our system allows for SNR improvements as more channels are added. Second, while the number of channels increases, noise due to dark counts remains constant because each detector only contributes counts when the pattern is assigned to that channel. In other words, the dark counts from all detectors only contributes when the time window is correlated to its energy channel, resulting in invariant total dark counts due to the same total time. This ability enables effective scaling of the system without sacrificing performance, making it highly suitable for practical remote sensing applications.

These advantages are clearly illustrated in Fig.~\ref{fig:Fig/Fig1}(a), when the detection range is set to 155~metres -- corresponding to our field trial distance with a transmission loss of 98.1~dB in bright sunlight -- the SNR is significantly enhanced under high background noise levels (e.g. $3\times 10^7$~Hz detected photon counts cross 100 Single-Photon Avalanche Diode (SPADs)) by increasing the number of energy channels from 1 to 100. 
Figure ~\ref{fig:Fig/Fig1}(b) depicts the increased loss during rainy conditions, demonstrating a more pronounced improvement under high background noise levels. This underscores the robustness of the correlation system in real-world scenarios, highlighting its potential for deployment in environments with elevated ambient light levels.

\bibliography{mylibrary}